\documentclass[runningheads]{llncs}
\usepackage{graphicx}
\usepackage{amsmath}
\usepackage{cite}  
\usepackage{graphicx}
\usepackage{makecell}
\usepackage{amsfonts}
\usepackage[caption=false]{subfig}
\usepackage{multirow}
\usepackage{amssymb}
\newcommand{\corrauth}{\textsuperscript{(\Letter)}}

\usepackage{lipsum}
\usepackage{mwe}
\usepackage[dvipsnames]{xcolor}
\usepackage{multirow}
\usepackage{cite}
\usepackage{amsmath}
\usepackage{tabu}
\usepackage{subfig}
\usepackage[misc]{ifsym}

\usepackage{booktabs}

\usepackage{color}

\newcommand{\etal}{\emph{et~al.}}

\begin{document}
\title{End-to-end Ultrasound Frame to Volume Registration}
%
%

\author{Hengtao Guo\inst{1} \and
Xuanang Xu\inst{1} \and
Sheng Xu\inst{2} \and
Bradford J. Wood\inst{2} \and\\
Pingkun Yan\inst{1}\corrauth}

%
\institute{Department of Biomedical Engineering and Center for Biotechnology and Interdisciplinary Studies, Rensselaer Polytechnic Institute, Troy, NY 12180, USA\\
\email{yanp2@rpi.edu}\\
\and
Center for Interventional Oncology, Radiology \& Imaging Sciences, National Institutes of Health, Bethesda, MD 20892, USA}

\maketitle              

\begin{abstract}
Fusing intra-operative 2D transrectal ultrasound (TRUS) image with pre-operative 3D magnetic resonance (MR) volume to guide prostate biopsy can significantly increase the yield. However, such a multimodal 2D/3D registration problem is very challenging due to several significant obstacles such as dimensional mismatch, large modal appearance difference, and heavy computational load. In this paper, we propose an end-to-end frame-to-volume registration network (FVR-Net), which can efficiently bridge the previous research gaps by aligning a 2D TRUS frame with a 3D TRUS volume without requiring hardware tracking. The proposed FVR-Net utilizes a dual-branch feature extraction module to extract the information from TRUS frame and volume to estimate transformation parameters. To achieve efficient training and inference, we introduce a differentiable 2D slice sampling module which allows gradients backpropagating from an unsupervised image similarity loss for content correspondence learning. Our experiments demonstrate the proposed method’s superior efficiency for real-time interventional guidance with highly competitive registration accuracy. Source code of this work is publicly available at \url{https://github.com/DIAL-RPI/FVR-Net}.


\end{abstract}

\section{Introduction}

Prostate cancer is a leading cause of cancer death for men in the United States \cite{siegel2019cancer}. Fusing transrectal ultrasound (TRUS) and magnetic resonance imaging (MRI) has been proven efficient for guiding targeted biopsies to more accurately diagnose the disease \cite{pinto2011magnetic,azizi2017detection}. Real-time 2D TRUS imaging is registered to a pre-operative 3D MRI volume for joint visualization during the fusion-guided procedures. Benefited by the time-efficient imaging of TRUS and high resolution of MRI, clinicians can locate the targeted lesions, thus increase the biopsy yield. The core of this technology is to register 2D TRUS images with a 3D MRI volume, which is a very challenging problem.


Existing fusion systems usually rely on external tracking devices to establish the registration \cite{xu2008real,natarajan2011clinical,bax2008mechanically,khallaghi20152d}. The workflow involves reconstructing a 3D TRUS volume from a sequence of tracked 2D TRUS video frames, which is then aligned with the pre-operative MRI volume through 3D-3D image registration. During the interventional guidance stage, a tracked 2D TRUS frame is mapped to the 3D TRUS volume and then transformed into the MRI image space for fusion. These tracking-based methods require a hardware setup, which induces additional cost and human effort.


Recent advances in deep learning (DL) based image registration and volume reconstruction have enabled new opportunities to shift the MRI/TRUS fusion paradigm. Hu \etal\cite{hu2018weakly} first proposed a weakly supervised method that uses landmark annotations as auxiliary information for training an end-to-end registration network. Haskins \etal\cite{haskins2019learning} developed a convolutional neural network (CNN) to learn the deep similarity metric between TRUS and MRI volume for the iterative registration. Guo \etal\cite{guo2020cmig} proposed a multi-stage registration framework that aligns a TRUS/MRI pair from coarse to fine. Deep learning has also been used for sensorless US volume reconstruction. Prevost \etal~\cite{prevost20183d} proposed to use a CNN for directly estimating the inter-frame motion between two 2D US frames, which enables sensorless US volume reconstruction. One recent work~\cite{guo2020sensorless} applies 3D CNN on a US video sub-sequence to better utilize the temporal context information for sensorless TRUS volume reconstruction. With the existing efforts, we are one step away from building a DL-based trackingless fusion system.

The objective of our work presented in this paper is to bridge the above research gap by developing a 2D TRUS image to 3D TRUS volume trackingless registration method. 2D/3D image registration is also often referred as slice-to-volume registration. Conventional approaches have tried to optimize the registration field according to an image matching criterion, which quantifies the alignment between the images and guides the optimization process. Classical matching criteria use pixel/voxel intensities to quantify the image similarity. For example, Wein \etal\cite{wein2007simulation} proposed a similarity measure named linear correlation of linear combination, which reveals the correspondence between simulated US images with MRI/Computed tomography (CT). However, the iterative optimization methods are time-consuming, typically taking seconds or more to register a single pair of images, making them unsuitable for interventional guidance. Another group of 2D/3D registration methods aims to align a 2D projective image with a 3D volume, which typically applies to 2D X-ray with 3D CT volume registration. For example, Miao \etal\cite{miao2016cnn, miao2018dilated} proposed to use CNN to directly predict the transformation parameters to perform 3D model to 2D X-ray registration. However, such projective 2D/3D registration is entirely different from our targeted application and thus not applicable.
In this paper, we propose a novel end-to-end frame-to-volume registration network (FVR-Net) to bridge the gap in real-time TRUS/MRI fusion for guiding prostate biopsy. The underlying framework takes one real-time 2D TRUS frame and one reconstructed 3D TRUS volume as input to estimate the transformation parameters that best align these two images. We proposed a dual-branch balanced feature extraction network, which makes the model sensitive to both the frame and volume information. Besides,  we introduce an auxiliary image similarity loss for end-to-end training which can significantly reduce the registration error. The experiments demonstrate that using CNN for FVR problems is highly promising, achieving competing performance to the conventional iterative registration methods while running tens of times faster.


\begin{figure}[t]
	\centering
	\includegraphics[width=\textwidth]{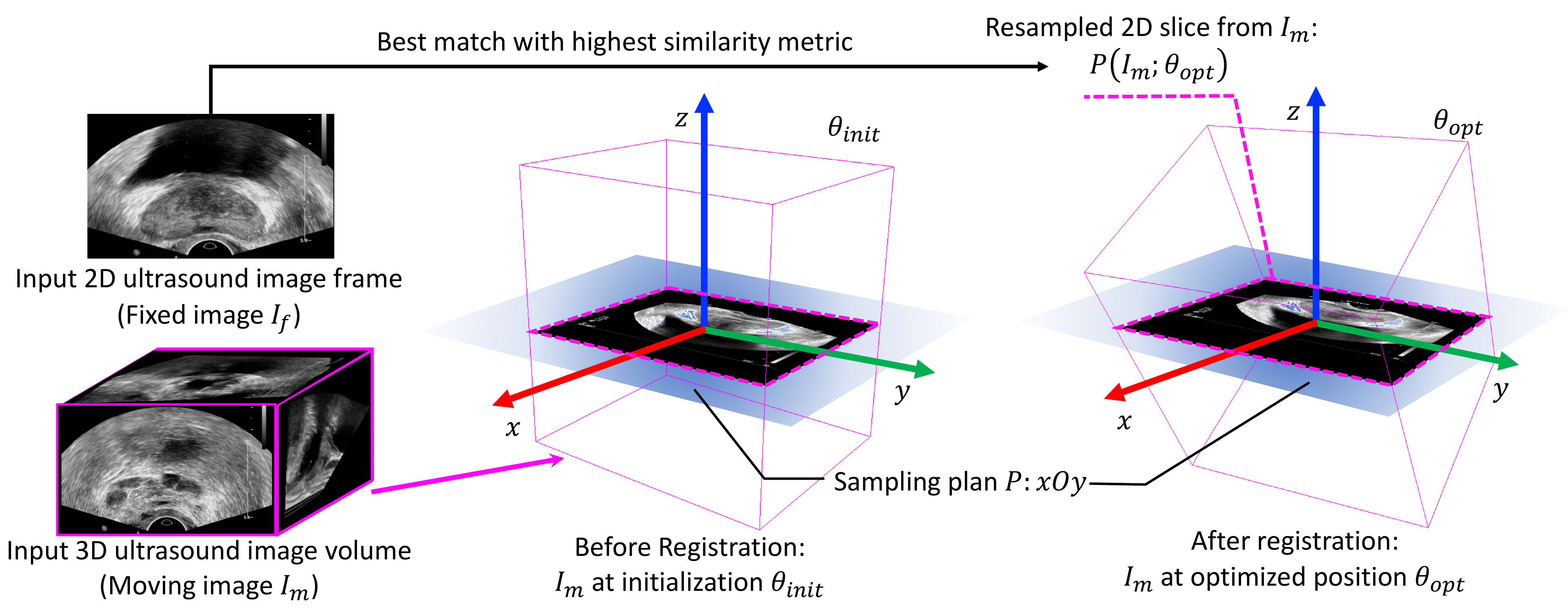}
	\caption{Illustration of the rigid frame-to-volume registration in this work. The pink cube represents the boundaries of the input 3D ultrasound image volume.}
	\label{fig:svr}
\end{figure}

\section{Problem Definition}
\label{sec:materials}


In this section, we give a formal definition of the mono-modal 2D TRUS frame to 3D TRUS volume registration problem. Fig.~\ref{fig:svr} illustrates the overall implementation workflow. Given a 2D TRUS frame $I_f$ (fixed image) and a 3D TRUS volume $I_m$ (moving image) as input, we seek a mapping function $\theta_{opt}$ through minimizing the following objective function:
\begin{equation}
    \theta_{opt}=\arg\min_{\theta}Sim(I_f, P(I_m;\theta)),
\label{eq:obj}
\end{equation}
where $P(I_m;\theta)$ denotes the slice extracted from $I_m$ and specified by the transformation $\theta$ and the sampling plane $P$, which is permanently set to be the $xOy$ plane. $Sim()$ is the matching criterion, which quantifies the image similarity between the 2D frame $I_f$ and the resampled slice $P(I_m;\theta)$. By default, for a volume transformed by an identical transformation $\theta_{init}$ as the initialization, the volume's center is placed at the coordinate system's origin. The goal of the frame-to-volume registration is to find the optimal transformation parameters $\theta_{opt}$, such that the resampled slice $P(I_m;\theta_{opt})$ has the highest image similarity as the 2D input frame $I_f$.




\section{Method}
\label{sec:method}

\begin{figure}[t]
	\centering
	\includegraphics[width=\textwidth]{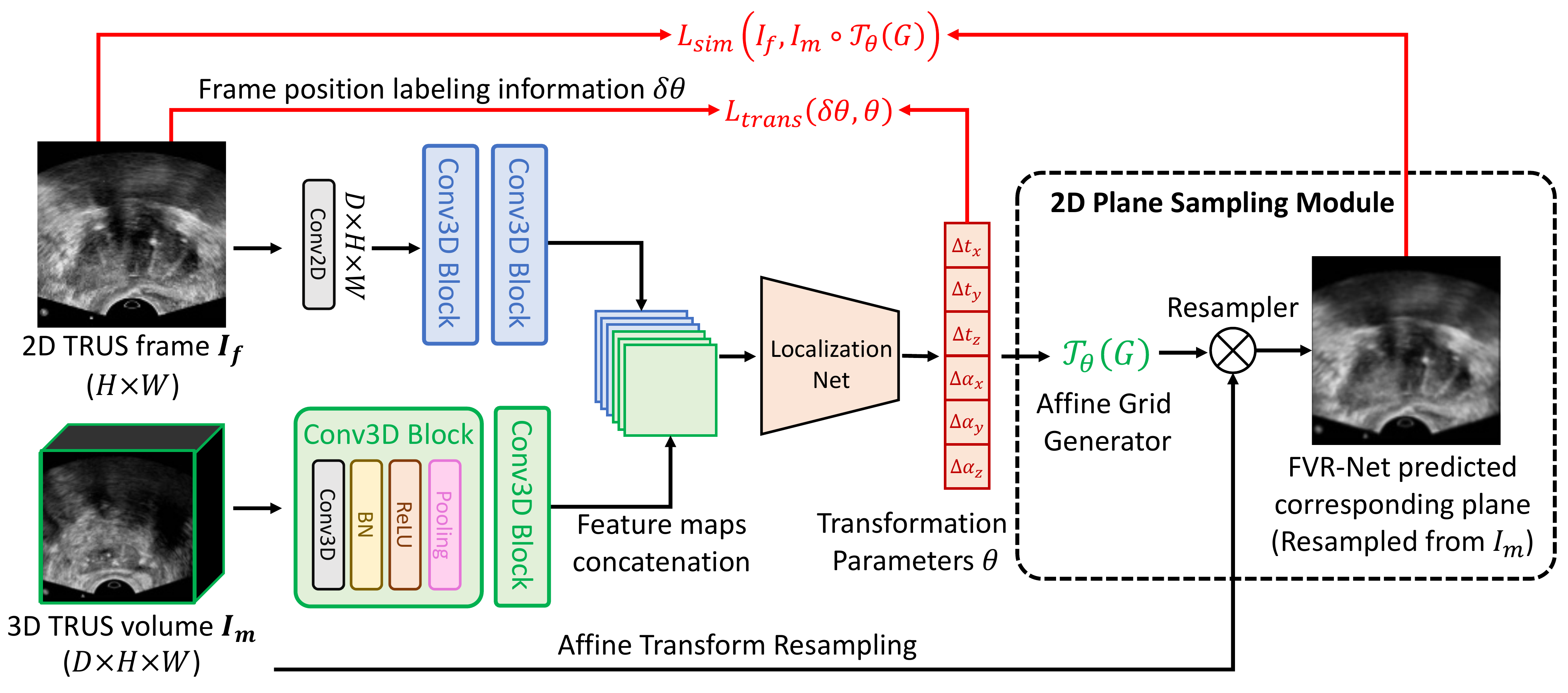}
	\caption{Overall network structure of the proposed FVR-Net.} 
	\label{fig:net}
\end{figure}

This section presents the key components of the proposed method.
Fig.~\ref{fig:net} depicts the proposed end-to-end frame-to-volume registration network (FVR-Net). The FVR-Net takes an 2D TRUS frame and a 3D volume as input for estimating the transformation $T(\theta)$, such that the transformed volumetric image and the sampling plane's cross-sectional area $P(I_m;\theta)$ has the highest image similarity with the 2D TRUS image frame $I_f$. We found the rigid registration can best suit in our application without loss of generality. Thus, the FVR-Net's output $\theta$ contains 6 degrees of freedom, i.e. $\theta={\left \{t_x,t_y,t_z,\alpha_x,\alpha_y,\alpha_z \right \}}$, including the translations and rotations along the three axes.

\subsection{End-to-end Slice-to-Volume Registration}

In this end-to-end FVR framework, we define the real-time 2D transrectal ultrasound image frame as the fixed image $I_f$ ($H \times W$), and the reconstructed 3D TRUS image subvolume as the moving image $I_m$ ($D\times H\times W$).

\noindent\textbf{Dual-branch Balanced Feature Extraction}
The dimension gap between 2D and 3D images is a major obstacle for the registration performance. Directly concatenating these two inputs together (early-fusion) can make the network overwhelmed by the volumetric information while completely ignoring the 2D image contents. Instead, to balance the data information, we designed a dual-branch network structure to extract the image features from the frame and volume independently and then concatenate them in a late-fusion fashion. 

In the frame branch, we first use a 2D convolutional layer to extract the low-level features for the input frame and extend the channel number to $D$, such that the feature map's size matches the input volume's size, thus achieving the data information balance. From this point, each branch is followed by two 3D convolutional blocks with the same hyper-parameters to maintain the identical feature map size. The extracted feature maps from both branches are concatenated together along the depth dimension and then serve as the input to the localization-net for estimating the transformation parameters $\theta$. A supervised mean squared error (MSE) loss can be computed, which directly uses ground-truth labeling information to constraint the $\theta$ estimation:
\begin{equation}
    L_{trans}=\frac{1}{N}\sum_{n=1}^{N}\left \| \delta\theta_{n}-\theta_{n} \right \|_{2}, 
\label{eq:gt}
\end{equation}
where $N$ denotes the total number of 2D/3D sample pairs within one training epoch, and $\delta\theta$ denotes the transformation parameters label. The localization-net uses ResNext\cite{xie2017aggregated} as the backbone structure. The proposed dual-branch feature extraction can ensure that the most representative image features can be learned from the images with different dimensions.

\noindent\textbf{Differentiable 2D Slice Sampling}
Inspired by the spatial transformer network (STN)\cite{jaderberg2015spatial}, we designed a 2D slice sampling module that can introduce a new unsupervised image similarity loss for stabilizing the training process. This module has two components: the affine grid generator and the resampler. Our customized affine grid generator takes the estimated parameters $\theta$ as input and generates a transformed resampling grid $T_\theta$ (G), which has the same size as the moving image $m$. By applying bilinear interpolation at each point location defined by the sampling grid, the resampler can get the intensity at a particular pixel in the wrapped image. Thus, a target 2D slice $P(I_m;\theta)$, noted as $I_m\circ T_\theta(G)$ in Fig.\ref{fig:net}, can be sampled from the 3D input volume $I_m$ transformed by the estimated parameters $\theta$. This FVR-Net predicted target slice denotes the results from the 2D/3D registration framework and should ideally contain the same information as the input 2D frame. 

Through the partial derivative, the loss gradients can be backpropagated to the sampling grid coordinates, and furthermore, to the affine transformation parameters and the localization net. This makes the entire pipeline of the FVR-Net differentiable and can be easily trained in an end-to-end manner\cite{jaderberg2015spatial}. We further introduce an auxiliary image similarity loss, modified from Eq.~\ref{eq:obj}:
\begin{equation}
    L_{sim}=\frac{1}{N}\sum_{n=1}^{N}\left \| I_{f,n}-P(I_{m,n};\theta_{n}) \right \|_{2}, 
\label{eq:img}
\end{equation}
Theoretically, the FVR-Net can be trained in an unsupervised fashion by using the $L_{sim}$ alone. However, in practice, we found this yielding an unstable training process. When $L_{trans}$ and $L_{sim}$ are used together to update the network's parameters, the FVR-Net shows the most robust registration performance. 

\subsection{Implementation Details}

To help the network focus on the prostate-relevant features, 2D input frames are center-cropped with a window size of $128\times128$ pixels. Ideally, the 3D volume should be kept intact such that the volume can always contain a full view of the 2D frame for registration. However, such an implementation encounters two difficulties. (1) The reconstructed volumes are in different sizes, making it difficult to design a network structure with fixed input and output sizes. (2) The volumes’ average size is too big, making the searching space too big for the network to find the optimal transformation.

We propose a data sampling strategy to solve the above issues, which also serves as an augmentation method for network training. By referring to the clinical application, we propose to sample a smaller subvolume as the searching space based on a random initial transformation $\theta_{init}$ within a manually defined range $R$. For example, if we are looking for the $n$th frame $f_n$'s position $\theta_n$ during the registration, one neighboring frame within the range of $[n-R, n+R]$ is randomly chosen as the initial reference frame $f_{init}$, which has the position of $\theta_{init}$. A subvolume size of $128\times128\times32$ is cropped at the center of $f_{init}$, which serves as the 3D volume input to the network. In our experiments, we set the frame range $R=10$ to ensure that the subvolume around the initial frame $f_{init}$ contains the target frame $f_n$. Taking this subvolume as the input, the network is trained to estimate the relative transformation parameters $\delta\theta$ which is the difference between $\theta_n$ and $\theta_{init}$, calculated through matrix manipulation.

\section{Experiments and Results}
\label{sec:results}
\subsection{Datasets and Experimental Setting}
All the ultrasound volumes used in this work were collected from clinical studies using an EM-tracking enabled fusion system. The dataset contains 619 TRUS volumes reconstructed from tracked TRUS frames acquired by a Philips iU22 scanner, all from different subjects. The dataset is further divided into 488, 65, 66 cases for training, testing, and validation. Each TRUS frame has an associated positioning transformation matrix $M$, describing the spatial relationship between the frame and the reconstructed volume. We use this information as the ground truth label for network training and validation. Our network was trained for 150 epochs with batch size $K=24$ using Adam optimizer~\cite{kingma2014adam} with an initial learning rate of 5$\times 10^{-5}$, which decays by 0.9 after every five epochs. We implemented the FVR-Net using the Pytorch library~\cite{pytorch}.

\subsection{Results Evaluation}

Since DL-based projective 2D/3D registration does not apply well to our problem, we compare our method with conventional iterative image registration methods as shown in Table~\ref{tab:res}. The baseline methods use mean square error (MSE) or normalized cross-correlation (NCC) as the image similarity metric, optimized with gradient descent~\cite{klein2009adaptive} (GD) or Powell~\cite{fletcher1963rapidly} optimizer. For the random guess, the transformation parameters were randomly sampled from the training set's statistics. 

\begin{table}[]
\caption{Performance comparison of our FVR-Net and baseline methods.}

\label{tab:res}
\begin{tabular}{l|c|c|ccccccc|c}
\toprule
\multicolumn{1}{l|}{\multirow{2}{*}{\textbf{Method}}} & \multirow{2}{*}{\begin{tabular}[c]{@{}c@{}}\textbf{DistErr}\\ (mm)\end{tabular}} & \multirow{2}{*}{\begin{tabular}[c]{@{}c@{}}\textbf{ImgSim}\\ (NCC)\end{tabular}} & \multicolumn{6}{c}{\textbf{Correlation}}            &      & \multirow{2}{*}{\begin{tabular}[c]{@{}c@{}}\textbf{RunTime}\\ (s)\end{tabular}} \\
\multicolumn{1}{l|}{}                  &                                                                         &                                                                         & tX    & tY   & tZ   & aX    & aY   & aZ    & Mean &                                                                        \\ \hline
Random Guess                           & 5.86                                                                    & 0.55                                                                    & -0.01 & 0.05 & 0.03 & -0.08 & 0.05 & -0.06 & 0.00 & -                                                                      \\
NCC + GD                               & 3.43                                                                    & 0.91                                                                    & 0.70  & 0.70 & 0.94 & 0.66  & 0.29 & 0.59  & 0.64 & 3.98                                                                   \\
MSE + GD                                & 3.20                                                                    & 0.90                                                                    & 0.60  & 0.66 & 0.94 & 0.64  & 0.21 & 0.55  & 0.60 & 3.64                                                                   \\
NCC + Powell                           & 3.02                                                                    & 0.91                                                                    & 0.43  & 0.86 & 0.89 & 0.80  & 0.13 & 0.51  & 0.60 & 7.60                                                                   \\
MSE + Powell                            & 2.85                                                                    & 0.92                                                                    & 0.64  & 0.90 & 0.99 & 0.79  & 0.36 & 0.63  & \textbf{0.72} & 5.85                                                                   \\
FVR-Net (Best)                         & \textbf{2.73}                                                                  & \textbf{0.92}                                                                    & 0.69  & 0.88 & 0.96 & 0.96  & 0.17 & 0.08  & 0.62 & \textbf{0.07}                                                                   \\ \bottomrule
\end{tabular}
\end{table}

\begin{figure}[t]
	\centering
	\includegraphics[width=\textwidth]{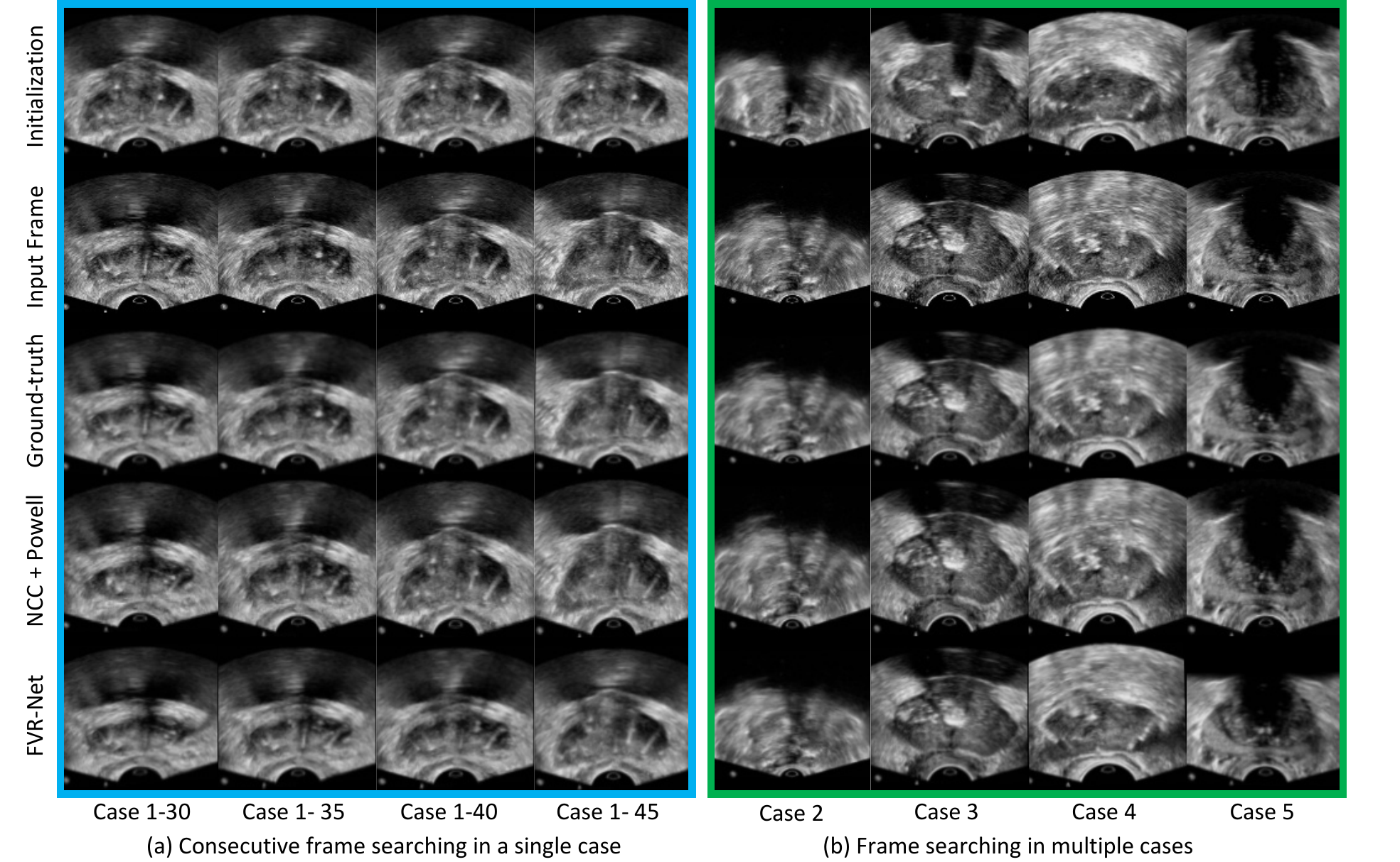}
	\caption{Sample results of frame-to-volume registration. The 1st row shows the frames sampled using the initial transformation; the 2nd row includes the input frames to be registered; the 3rd row is the target slice sampled from a 3D volume using the ground-truth transformation; the 4th and 5th rows are the results produced by the baseline registration method and our proposed FVR-Net.} 
	\label{fig:case}
\end{figure}

For evaluation, the distance error (DistErr) denotes the average distance in millimeters between the groundtruth-sampled slice and the predicted-sampled slice's corresponding corner points. The image similarity score (ImgSim) shows the quality assessment of the registration results. We also report the correlation coefficient between the groundtruth $\delta\theta$ and the estimated $\theta$. Finally, the running time denotes the average time cost for registering a single pair of images. The average initialization error is 7.68 mm in the test set. As shown in Table \ref{tab:res}, the iterative methods using MSE as the similarity metric can consistently outperform those using NCC. On the other hand, using Powell as the optimizer, the distance error is significantly reduced, but the running time also increased rapidly. By applying our best FVR-Net, a competitive performance (2.73mm) has been achieved compared to the best iterative registration (2.85mm). Although there is no significant difference in the distance error, our running time is more than 80 times faster than the traditional iterative registration. Benefitted from deep CNN’s high computation efficiency, the proposed FVR-Net can perform TRUS frame-to-volume registration in approximately real-time speed. Fig.\ref{fig:case} (a) shows the results of searching for consecutive frames within one volume. Given the same input subvolume (specified by the same initialization frame), our FVR-Net can search for different frames within this subvolume. The FVR-Net results are similar to the groundtruth sampling (3rd) row. 


\begin{table}[t]
\label{tab:ablation}
\caption{Ablation studies of FVR-Net using ResNext-150 unless specifically noted.}
\begin{tabular}{l|c|c|ccccccc|c}
\toprule
\multirow{2}{*}{\textbf{Method}}                                              & \multirow{2}{*}{\begin{tabular}[c]{@{}c@{}}\textbf{DistErr}\\ (mm)\end{tabular}} & \multirow{2}{*}{\begin{tabular}[c]{@{}c@{}}\textbf{ImgSim}\\ (NCC)\end{tabular}} & \multicolumn{6}{c}{\textbf{Correlation}}                                                               & \multicolumn{1}{l|}{} & \multirow{2}{*}{\begin{tabular}[c]{@{}c@{}}\textbf{Runtime}\\ (s)\end{tabular}} \\
 &                                                                         &                                                                         & tX            & tY            & tZ            & aX            & aY            & aZ            & Mean                  &                                                                        \\ \midrule
ResNext-50                                                              & 4.34                                                                    & 0.72                                                                    & 0.69          & 0.76          & 0.94          & 0.95          & -0.06         & 0.05          & 0.55                  & \textbf{0.03}                                                          \\
ResNext-101 
& 3.35                                                                    & 0.85                                                                    & 0.66          & 0.68          & 0.87          & 0.88          & 0.10          & -0.01         & 0.53                  & 0.04                                                                   \\
$L_{trans}$                                                   & 3.17                                                                    & 0.90                                                                    & 0.48          & 0.56          & 0.91          & 0.91          & 0.01          & \textbf{0.14} & 0.50                  & 0.07                                                                   \\
$L_{sim}$                                                     & 4.86                                                                    & 0.61                                                                    & 0.01          & 0.05          & 0.03          & -0.08         & 0.05          & -0.06         & 0.00                  & 0.07                                                                   \\
EF                                                         & 5.68                                                                    & 0.60                                                                    & 0.14          & 0.06          & 0.07          & 0.14          & 0.01          & 0.04          & 0.03                  & 0.06                                                                   \\
ULF                                                        & 5.93                                                                    & 0.62                                                                    & 0.18          & 0.07          & 0.08          & -0.16         & 0.08          & -0.02         & 0.04                  & 0.06                                                                   \\
$L_{trans}+L_{sim}$
& \textbf{2.73}                                                           & \textbf{0.92}                                                           & \textbf{0.69} & \textbf{0.88} & \textbf{0.96} & \textbf{0.96} & \textbf{0.17} & 0.08          & \textbf{0.62}         & 0.07             \\ \bottomrule
\end{tabular}
\end{table}

\noindent\textbf{Ablation Study}
To determine whether the superior result of our network attributes to the novel (a) dual-branch balanced feature extraction and (b) auxiliary image similarity loss, we trained our FVR-Net with multiple settings, as shown in Table~\ref{tab:ablation}. We tried different architectures of the localization net and found ResNext-150 produces the smallest distance error. By directly concatenating the 2D frame and 3D volume together as input, the early fusion (EF) network produces meaningless results, indicating the network learns little information for the transformation parameters estimation. A similar phenomenon happens to the unbalanced late fusion (ULF), where the frame's feature maps are still overwhelmed by the volume feature maps' large size. Applying our dual-branch balanced feature extraction, the resultant transformation parameters show a high correlation to the groundtruth. When the FVR-Net is trained in an unsupervised way using $L_{sim}$ alone, the results show little correlation to the groundtruth. While combining two loss functions, the distance error drops significantly from 3.17mm to 2.73mm, demonstrating the effectiveness of the auxiliary image similarity loss’s effectiveness.

\section{Conclusions}
\label{sec:conclusions}
This paper has introduced a novel end-to-end TRUS frame-to-volume registration network which instantly register a single 2D TRUS frame with a reconstructed 3D TRUS volume. The experimental results demonstrate our work's competitive performance and superior registration speed comparing to the conventional iterative registration. The ablation study suggests that the dual-branch balanced feature extraction and auxiliary image similarity loss can significantly reduce the registration error. 
Once combined with off-the-shelf 3D TRUS-MRI registration methods\cite{guo2020cmig,haskins2019learning}, the proposed FVR-Net would be able to bridge the research gap in real-time 2D-TRUS and 3D-MRI fusion for guiding the prostate biopsy. We will systematically study the entire workflow in our future work and include more ablation studies for various experimental settings.

\section{Acknowledgements}
\label{Acknowledgements}
This work was partially supported by National Institute of Biomedical Imaging and Bioengineering (NIBIB) of the National Institutes of Health (NIH) under awards R21EB028001 and R01EB027898, and through an NIH Bench-to-Bedside award made possible by the National Cancer Institute.

%
%
%
\bibliographystyle{splncs04} 
\bibliography{references}

\begin{thebibliography}{10}
\providecommand{\url}[1]{\texttt{#1}}
\providecommand{\urlprefix}{URL }
\providecommand{\doi}[1]{https://doi.org/#1}

\bibitem{azizi2017detection}
Azizi, S., Bayat, S., Yan, P., Tahmasebi, A., Nir, G., Kwak, J.T., Xu, S.,
  Wilson, S., Iczkowski, K.A., Lucia, M.S., et~al.: Detection and grading of
  prostate cancer using temporal enhanced ultrasound: combining deep neural
  networks and tissue mimicking simulations. International journal of computer
  assisted radiology and surgery  \textbf{12}(8),  1293--1305 (2017)

\bibitem{bax2008mechanically}
Bax, J., Cool, D., Gardi, L., Knight, K., Smith, D., Montreuil, J., Sherebrin,
  S., Romagnoli, C., Fenster, A.: Mechanically assisted 3d ultrasound guided
  prostate biopsy system. Medical physics  \textbf{35}(12),  5397--5410 (2008)

\bibitem{fletcher1963rapidly}
Fletcher, R., Powell, M.J.: A rapidly convergent descent method for
  minimization. The computer journal  \textbf{6}(2),  163--168 (1963)

\bibitem{guo2020cmig}
Guo, H., Kruger, M., Xu, S., Wood, B.J., Yan, P.: Deep adaptive registration of
  multi-modal prostate images. Computerized Medical Imaging and Graphics
  \textbf{84},  101769 (2020)

\bibitem{guo2020sensorless}
Guo, H., Xu, S., Wood, B., Yan, P.: Sensorless freehand {3D} ultrasound
  reconstruction via deep contextual learning. In: International Conference on
  MICCAI. pp. 463--472. Springer (2020)

\bibitem{haskins2019learning}
Haskins, G., Kruecker, J., Kruger, U., Xu, S., Pinto, P.A., Wood, B.J., Yan,
  P.: Learning deep similarity metric for 3d mr--trus image registration.
  International journal of computer assisted radiology and surgery
  \textbf{14}(3),  417--425 (2019)

\bibitem{hu2018weakly}
Hu, Y., Modat, M., Gibson, E., Li, W., Ghavami, N., Bonmati, E., Wang, G.,
  Bandula, S., Moore, C.M., Emberton, M., et~al.: Weakly-supervised
  convolutional neural networks for multimodal image registration. Medical
  image analysis  \textbf{49},  1--13 (2018)

\bibitem{jaderberg2015spatial}
Jaderberg, M., Simonyan, K., Zisserman, A., Kavukcuoglu, K.: Spatial
  transformer networks. arXiv preprint arXiv:1506.02025  (2015)

\bibitem{khallaghi20152d}
Khallaghi, S., S{\'a}nchez, C.A., Nouranian, S., Sojoudi, S., Chang, S., Abdi,
  H., Machan, L., Harris, A., Black, P., Gleave, M., et~al.: A 2d-3d
  registration framework for freehand trus-guided prostate biopsy. In:
  International Conference on Medical Image Computing and Computer-Assisted
  Intervention. pp. 272--279. Springer (2015)

\bibitem{kingma2014adam}
Kingma, D.P., Ba, J.: Adam: A method for stochastic optimization. arXiv
  preprint arXiv:1412.6980  (2014)

\bibitem{klein2009adaptive}
Klein, S., Pluim, J.P., Staring, M., Viergever, M.A.: Adaptive stochastic
  gradient descent optimisation for image registration. International journal
  of computer vision  \textbf{81}(3), ~227 (2009)

\bibitem{miao2018dilated}
Miao, S., Piat, S., Fischer, P., Tuysuzoglu, A., Mewes, P., Mansi, T., Liao,
  R.: Dilated fcn for multi-agent {2D/3D} medical image registration. In:
  Proceedings of the AAAI Conference on Artificial Intelligence. pp. 4694--4701
  (2018)

\bibitem{miao2016cnn}
Miao, S., Wang, Z.J., Liao, R.: A cnn regression approach for real-time 2d/3d
  registration. IEEE transactions on medical imaging  \textbf{35}(5),
  1352--1363 (2016)

\bibitem{natarajan2011clinical}
Natarajan, S., Marks, L.S., Margolis, D.J., Huang, J., Macairan, M.L., Lieu,
  P., Fenster, A.: Clinical application of a 3d ultrasound-guided prostate
  biopsy system. In: Urologic oncology: seminars and original investigations.
  vol.~29, pp. 334--342. Elsevier (2011)

\bibitem{pytorch}
Paszke, A., Gross, S., Chintala, S., Chanan, G., Yang, E., DeVito, Z., Lin, Z.,
  Desmaison, A., Antiga, L., Lerer, A.: Automatic differentiation in pytorch.
  In: NIPS 2017 Workshop Autodiff (2017)

\bibitem{pinto2011magnetic}
Pinto, P.A., Chung, P.H., Rastinehad, A.R., Baccala, A.A., Kruecker, J.,
  Benjamin, C.J., Xu, S., Yan, P., Kadoury, S., Chua, C., et~al.: Magnetic
  resonance imaging/ultrasound fusion guided prostate biopsy improves cancer
  detection following transrectal ultrasound biopsy and correlates with
  multiparametric magnetic resonance imaging. The Journal of urology
  \textbf{186}(4),  1281--1285 (2011)

\bibitem{prevost20183d}
Prevost, R., Salehi, M., Jagoda, S., Kumar, N., Sprung, J., Ladikos, A., Bauer,
  R., Zettinig, O., Wein, W.: {3D} freehand ultrasound without external
  tracking using deep learning. Medical image analysis  \textbf{48},  187--202
  (2018)

\bibitem{siegel2019cancer}
Siegel, R.L., Miller, K.D., Jemal, A.: Cancer statistics, 2019. CA: a cancer
  journal for clinicians  \textbf{69}(1),  7--34 (2019)

\bibitem{wein2007simulation}
Wein, W., Khamene, A., Clevert, D.A., Kutter, O., Navab, N.: Simulation and
  fully automatic multimodal registration of medical ultrasound. In:
  International Conference on Medical Image Computing and Computer-Assisted
  Intervention. pp. 136--143. Springer (2007)

\bibitem{xie2017aggregated}
Xie, S., Girshick, R., Doll{\'a}r, P., Tu, Z., He, K.: Aggregated residual
  transformations for deep neural networks. In: Proceedings of the IEEE
  conference on computer vision and pattern recognition. pp. 1492--1500 (2017)

\bibitem{xu2008real}
Xu, S., Kruecker, J., Turkbey, B., Glossop, N., Singh, A.K., Choyke, P., Pinto,
  P., Wood, B.J.: Real-time mri-trus fusion for guidance of targeted prostate
  biopsies. Computer Aided Surgery  \textbf{13}(5),  255--264 (2008)

\end{thebibliography}

\end{document}